\documentclass[a4paper,10pt]{article}

\usepackage{amssymb}
\usepackage{amsfonts}
\usepackage{amsmath}
\usepackage{subfigure}
\usepackage{algorithm}
\usepackage[noend]{algorithmic}
\usepackage{amsmath,amssymb,cite}
\usepackage{epsfig}
\usepackage{pst-node,pstricks}
\usepackage{enumerate}

\usepackage{graphicx}
\usepackage{psfrag}
\usepackage[nohead, top=2.9cm, bottom=3.2cm, left=2.9cm, right=3.1cm]{geometry}

\newtheorem{definition}{Definition}[section]

\def\q5uad{\quad\quad\quad\quad\quad}

\title{Malicious Software Detection and Classification utilizing Temporal-Graphs of System-call Group Relations \vspace{0.2cm}}

\author{Anna~Mpanti \ \ Stavros~D.~Nikolopoulos \ \ Iosif~Polenakis }

\date{}

\begin{document}

\maketitle

\vspace{-0.5cm}

\centerline{\it Department of Computer Science \& Engineering}

\centerline{\it University of Ioannina}

\centerline{\it GR-45110 Ioannina, Greece}

\centerline{\tt \{ampanti,stavros,ipolenak\}@cs.uoi.gr}


\vskip 0.3in

\begin{abstract}
\noindent In this work we propose a graph-based model that, utilizing relations between groups of System-calls, distinguishes malicious from benign software samples and classifies the detected malicious samples to one of a set of known malware families. More precisely, given a System-call Dependency Graph (ScDG) that depicts the malware's behavior, we first transform it to a more abstract representation, utilizing the indexing of System-calls to a set of groups of similar functionality, constructing thus an abstract and mutation-tolerant graph that we call Group Relation Graph (GrG); then, we construct another graph representation, which we call Coverage Graph (CvG), that depicts the dominating relations between the nodes of a GrG graph. Based on the research so far in the field, we pointed out that behavior-based graph representations had not leveraged the aspect of the temporal evolution of the graph. Hence, the novelty of our work is that, preserving the initial representations of GrG and CvG graphs, we focus on augmenting the potentials of theses graphs by adding further features that enhance its abilities on detecting and further classifying to a known malware family an unknown malware sample. To that end, we construct periodical instances of the graph that represent its temporal evolution concerning its structural modifications, creating another graph representation that we call Temporal Graphs. In this paper, we present the theoretical background behind our approach, discuss the current technological status on malware detection and classification and demonstrate the overall architecture of our proposed detection and classification model alongside with its underlying main principles and its structural key-components.

\vspace*{0.1in}
\end{abstract}

\vspace*{0.4in}
\section {Introduction}
\label{sec:Introduction}
\vspace*{0.1in}

\noindent Malware or malicious software is a software type intended to cause harm to end point computers, systems or networks \cite{SiHo}. In this work we design and propose a graph based model that develops an algorithmic technic for malware detection and classification. Our method is applied on unknown software samples in order to detect whether they are malicious or not, and further classify them to one of a set of known malware families (i.e., set of malicious mawlare samples with similar functionality), as they have been developed by various antivirus software vendors.

\vspace*{0.1in}
\subsection{Malware and Mutations}

\vspace*{0.05in}
\noindent On the contrary part of our scientific field, malware authors, have developed and deployed various techniques in order to avoid the traditional byte-level signature based detection methods. Since such detection methods appear to be significantly fragile against even the least (i.e., bit-level) mutation of the initial subject (i.e., ancestor malware sample), they mutate their software products (malware) creating structurally different but functionally similar copies of them.  Except from the mutation methods that leverage one, or more, levels of encryption, there also exist more advanced mutation methods. Some of the most applicable malware mutations are the oligomorphism which is achieved through obfuscation techniques, the polymorphism where the code is modified through encryption techniques and the metamorphism, in which multiple structurally different copies of a malware sample are generated.

More precisely an {\it oligomorphic} or {\it semi-polymorphic} malware, is a specific category of obfuscated malware disposing an encryption/decryption module for multi-layer encryption in order to avoid decryption body detection. On the other hand, a {\it polymorphic} malware can create an endless number of new decryptors that use different encryption methods to encrypt the body of the malware \cite{SzFe}. As referred in \cite{RaMaSu}, the main principal is to modify the appearance of the code constantly across the copies. Finally, a {\it metamorphic} malware changes its structure while keeps its functionality each time it replicates itself \cite{MuMa}. Polymorphic and metamorphic malware is the hardest type of malware to detect, since are able to mutate in an infinite number of functionally equivalent copies of themselves, and thus there is not a constant signature for virus scanning \cite{MuMa}.

Hence, while the main functionality of a malware sample remains immutable during its mutations, malware samples can be merged into groups of malware samples with common functionality, the so called malware families. So, in this work we developed an algorithmic technic that not only detects if a program is malicious or not, but additionally, given a malicious software it can decide the malware family that it belongs to.

\vspace*{0.1in}
\subsection{Protection against Malicious Software}

\vspace*{0.05in}
\noindent Since malicious software poses a major threat, several protection approaches have been proposed and implemented in order to eliminate such threats. The main corpus of the defense line is mainly developed over three axes, namely malware analysis, malware detection and malware classification:

\vspace*{0.05in}
\noindent \textbf{Malware Analysis.} Malware analysis \cite{BaMoKrKi} is the process of determining the purpose and the functionality or, abstractly, the behavior of a given malicious code. Such a process is a necessary prerequisite in order to develop efficient and effective detection and also classification methods, and is mainly divided into two main categories, namely {\it Static} and {\it Dynamic} analysis \cite{SiHo}.

\vspace*{0.05in}
\noindent \textbf{Malware Detection.} The term \textit{malware detection} referrers to the process of determining whether a given program $P$ is malicious or benign according to an  {\it a priori knowledge} \cite{AlLaVeWa,ChJhSeSoBr,MaHi}. Specifically with the term {\it a priori knowledge} we are referred to something that is known to be malicious or a characteristic that owned by something that is malicious, at a given time. However, an efficient malware detection is strongly related to {\it malware analysis}, during which, the analyst collects all the required information.

\vspace*{0.05in}
\noindent \textbf{Malware Classification.} The term \textit{malware classification} refers to the process of determining the malware family to which a particular malware sample, let $M$ belongs to. Malware classification is a quite important procedure, since the indexing of malware samples into families provides the ability to generalize detection signatures from sample level to family level. Through the indexing of a malware sample to a malware family, the construction of a new sample-specific signature is omitted, since the sample can be detected by the signature of its family.

\vspace*{0.1in}
\subsection{Our Approach}

\vspace*{0.05in}
\noindent  In our approach, we leverage the use of behavioral graph representations of software samples in order to distinguish if they are malicious or not and further classify them to a malware family. More precisely, given a representation (behavioral graph) of the behavior of a malicious sofware, in our case a System-call Dependency Graph - ScDG, constructed capturing the dependencies of the system-calls invoked during the execution of a software, we construct a directed edge-weighted graph, which we call Group Relation Graph - GrG, resulting from ScDG after grouping disjoint subsets of its vertices. Such graph abstraction has been proven \cite{NiPo2} that by generalizing graphs structure makes the detection and classification processes more resilient to known malware mutation procedures. Next we construct an additional graph representation Coverage Graph - CvG, that is a vertex-weighted undirected graph which results from GrGs after computing domination relation among its vertex set (regarding their degree and weight) when representing them on the Cartesian plane. Throughout these processes, over specific time intervals, we preserve instances of GrGs and CvGs creating hence {\tt Temporal Graphs} that depict their structural evolution over time, namely Group Relation Temporal Graph - GrTG and Coverage Temporal Graph - CvTG, respectively. Given a ScDG graph that represents a known malware sample and a ScDG graph that represent an unknown one, we utilize these instances (i.e., Temporal Graphs) over their corresponding GrG and CvG, produced both on each ScDG, in order to perform graph similarity towards the processes of malware detection and classification.

\vspace*{0.1in}
\subsection{Contribution}

\vspace*{0.05in}
\noindent In this work, we present our graph-based model for distinguishing graph representations referencing malicious software and further classifying them in sets of known malware families. Firstly, we discuss our proposed graph abstractions over ScDG representing the relations between system-call groups (i.e., Group Relation Graphs - GrG) and its corresponding graph that represents dominating relations over the nodes of GrG (i.e., Coverage Graph - CvG). Moreover, we present another graph representation that describes the temporal evolution on the structure of the aforementioned graph abstractions (i.e., GrG and CvG) Group Relation Temporal Graph - GrTG and Coverage Temporal Graph - CvTG, respectively, leveraging the temporal correlation between the structural modifications of two graphs, in order to be utilized on graph similarity. Furthermore, we demonstrate the development of an integrated framework that implements graph similarity approaches over the deployed GrTG and CvTG graphs in order to perform the malware detection and classification processes. Finally, we discuss the potentials of our approach, setting our further research landmarks for the extensions of our proposed model and conclude our work.

\vspace*{0.1in}
\subsection{Related Work}

\vspace*{0.05in}

In malware detection, there have been proposed similar models utilizing different non graph-based techniques like the one proposed by Alazab {\it et~al.}~\cite{AlLaVeWa}, who developed a fully automated system that disassembles and extracts API-call features from executables and then, using $n$-gram statistical analysis, is able to distinguish malicious from benign executables. The mean detection rate exhibited was 89.74\% with 9.72\% false-positives when used a Support Vector Machine (SVM) classifier by applying $n$-grams. In~\cite{YeDiTaDo}, Ye {\it et~al.} described an integrated system for malware detection based on API-sequences. This is also a different model from ours since the detection process is based on matching the API-sequences on OOA rules (i.e., Objective-Oriented Association) in order to decide the maliciousness or not of a test program. Finally, an important work of Christodorescu {\it et~al.}, presented in~\cite{ChJhSeSoBr}, proposes a malware detection algorithm, called $\mathcal{A}_{MD}$, based on instruction semantics. More precisely, templates of control flow graphs are built in order to demand their satisfiability when a program is malicious. Although their detection model exhibits better results than the ones produced by our model, since it exhibits 0 false-positives, it is a model based on static analysis and hence it would not be fair to compare two methods that operate on different objects. Kolbitch {\it et al.}~\cite{KoCoKrKiZhWa} proposed an effective and efficient approach for malware detection, based on behavioral graph matching by detecting string matches in system-call sequences, that is able to substitute the traditional anti-virus system at the end hosts. The main drawback of this approach is the fact that although no false-positives where exhibited, their detection rates are too low compared with other approaches. Luh and Tavolato~\cite{LuTa} presented one more detection algorithm based on behavioral graphs that distinguishes malicious from benign programs by grading the sample based on reports generated from monitoring tools. While the produced false-positives are very close to ours, the corresponding detection ratio is even lower.

In malware classification, there have been proposed other non graph-based malware classification models. Among them, a scalable automated approach for malware classification using pattern recognition algorithms and statistical methods, is presented by Islam {\it et al.} in \cite{IsTiBaVe}, utilizing the combination of static features extracted by function length and printable strings. While their evaluation results are very high(i.e., $98.8$\% classification accuracy), however it is worth mentioning the fact that their experiments include samples from $13$ malware families, while the classification accuracy of the model proposed in this paper has been evaluated over $48$ malware families. Hence, concerning the impact of philogeny among different malware families the comparative difference between the classification rates achieved by these two models is totally justified, while increasing the number of families in the training set increases the chances of misclassifications. Recently, Nataraj {\it et al.} \cite{NaKaJaMa} classify malware samples using image processing techniques. Visualizing as gray-scale images the malware binaries, they utilize the fact that,for many malware families, the images belonging to the same family appear very similar in layout and texture. Obviously the results are better than the ones produce by our model however they use at most $25$ malware families for their large scale experiments, where the impact of philogeny among different malware families is decreased the less different malware families in the training are. Finally, in \cite{NaYePoZh} Nataraj {\it et al.} utilize a static analysis technique called binary texture analysis in order to classify malicious binary samples into malware families. They achieve a $72\%$ rate of consistent classification when performing their evaluation on a data set of $60K$ to $685K$ samples comparing their labels with those provided by AV vendors, proving both the accuracy and the scalability of their model.

In the most recent literature, Makandar and Patrot \cite{MaPa} focus on detection and classification of the Trojan viruses using image processing techniques. In their proposed algorithm Gabor wavelet is used for key of feature extraction method and their experimental results are analyzed compared with two classifications such as KNN and SVM. In \cite{HaCh}, Hassen and Chan investigate a linear time function call graph (FCG) vector representation based on function clustering that has significant performance gains in addition to improved classification accuracy. They also show how this representation can enable using graph features together with other non-graph features. Recently, Sikora and Zelinka \cite{SiZe} investigate how behavior of malicious software can be connected with evolution and visualization of its spreading as the network. Their approach is based on hypothetical swarm virus and its dynamics of spread in PC and they show that its dynamics can be then modeled as the network structure and thus likely controlled and stopped, as their experiments suggest. Later, Souri and Hosseini \cite{SoHo} present a systematic and detailed survey of the malware detection mechanisms using data mining techniques. Additionally, it classifies the malware detection approaches in two main categories including signature-based methods and behavior-based detection. Based on the dependency graphs of malware samples, Ding et al. \cite{DiXiCheLi} propose an algorithm to extract the common behavior graph for each known malware, which is used to represent the behavioral features of a malware family. In addition, a graph matching algorithm that is based on the maximum weight subgraph is used to detect malicious code. In \cite{MuRaUp}, Mukesh et al. propose a machine learning based architecture to distinguish existing and recently developing malware by utilizing network and transport layer traffic features.

\newpage

\vspace*{0.1in}
\subsection{Road Map}

\vspace*{0.05in}
\noindent In Section~\ref{sec:Theoretical} we present the prerequisite theoretical background in order for the graph-based techniques for detection and classification to be developed, next in Section~\ref{sec:Components} we demonstrate the key components of our model, in Section~\ref{sec:Architecture} we discuss extensively the main principles and design aspects over the development of our detection and classification scheme concerning the two processes, where in Section~\ref{sec:Conclusion} we set our further research landmarks discussing the potentials and limitations of our proposed model concerning the processes of malware detection and classification, and we present our concluding remarks.

\vspace*{0.2in}
\section{Conceptual Framework}
\label{sec:Theoretical}
\vspace*{0.1in}

\noindent In this section we discuss the semantics behind the processes of malware analysis, detection and classification. We discuss the principles of the utilization of behavior-based approaches towards the deployment of resilient detection and classification techniques. Firstly we present the major process preceding the development of detection and classification methods, that is malware analysis, and next we depict the state-of-the art behavioral approaches applied on malware detection and classification.

\vspace*{0.1in}
\subsection{Analyzing Susceptible Samples}

\vspace*{0.05in}

\noindent The traditional signature-based malware detection, despite its fast real-time protection, is still not resilient against malware mutations. Robust detection techniques prerequisite the procedure of {\it malware analysis}, during which, the analyst collects all the required information, in order to be effective and efficient. The effectiveness of signature scanning, relying on pattern matching fails to detect new malware strains or mutated variants of existing ones \cite{DaDiViAuSt}.

The procedure of malware analysis is consisted by the collecting of valuable information concerning either static artifacts or generally behavioral patterns, that could characterize the maliciousness, or not, of a sample, being categorized to two main categories namely static analysis and dynamic analysis, respectively \cite{SeLeDo,SuQi}. In a more abstract level, in static analysis the specimen (i.e., test sample) is examined without its execution, performing the analysis on its source code, utilizing reverse engineering techniques when source code is unavailable, while on the other hand, in dynamic analysis an execution of the malware has to be performed in order to collect the required data, concerning the behavior of a program \cite{BaReSo}.

\vspace*{0.05in}
\noindent \textbf{Static Analysis.} Static analysis of software is performed over the programming artifacts and structural characteristics of a software sample \cite{DaFiGa}, without the need of its execution. The information obtained during static malware analysis may refer to opcode sequences, control flow graphs, etc. and can be used at will for malware detection \cite{DaDiViAuSt}. In static malware analysis, since the sample does not need to be executed can be surpasses by easily foiled by obfuscation and packing techniques (change the sequence of instructions or the signatures of malware), however its scalability consists one of its assets \cite{SeLeDo,LiPaLi}.

Several approaches have been deployed on the implementation of static malware analysis, including control-flow graphs, function call graph, machine learning techniques, support vector machines, similarity between API call sequences and opcode sequences, hidden Markov models, and principal component analysis \cite{DaDiViAuSt}.

\vspace*{0.05in}
\noindent \textbf{Dynamic Analysis.}
Dynamic malware analysis deals mostly with the extraction of behavioral features exhibited during the execution of a malicious software sample. Such behavioral features include among others: environmental artifacts, timing, process introspection, network artifacts, etc \cite{BuYe}. The behavioral features are mainly captured and depicted through API-calls sequences and system-calls dependencies \cite{DaDiViAuSt}.

For security reasons the whole execution takes place inside a virtual emulated environment (i.e., a Virtual Machine) \cite{BaReSo}, however the scalability of dynamic malware analysis may be reduced due to the demand of real time execution \cite{SeLeDo}. Moreover, despite that obfuscation techniques can easily be defeated through dynamic analysis the time needed for analysis is disproportionate to the rate of birth of mutated malware samples \cite{LiPaLi}. Hence, the need for automated dynamic analysis leaded to the development of integrated dynamic analysis systems that corporate visualized and supervision environments i.e., virtualized analysis systems, which among others include: emulators, hosted virtual machines, hypervisors, etc \cite{BuYe}.

A specific type of dynamic analysis, called {\it taint analysis} or DTA (stands for dynamic taint analysis) traces data flows in programs or systems during execution time. Briefly speaking, taint analysis distinguishes three elements namely {\it taint sources}, {\it taint sinks}, and {\it propagation rules}. Data flows are taint variables introduced by taint sources (i.e., the output parameters of system calls) and propagated according to the propagation rules to taint sinks (i.e., the input parameters of system calls)\cite{BaReSo}. However, through the literature, there have been proposed several techniques that combine characteristics from both static and dynamic approaches, synthesizing a hybrid analysis model \cite{DaDiViAuSt}.

\vspace*{0.1in}
\subsection{Detecting Malicious Behaviors}

\vspace*{0.05in}
\noindent A we mentioned previously, malicious software samples are intended to compromise the privacy, the confidentiality or the integrity of a system, of data or any other cyber-source constituting hence an intrusion. To this end, Intrusion Detection Systems, or, for short IDS, are deployed in order to monitor the execution of applications, the traffic of networks or whole systems, aiming on spotting malicious activity patterns \cite{AnBa}. The system supervision through an IDS can be performed through the application of {\it malware detection} techniques, that reference file comparisons against signatures of malicious software \cite{ElAnQuBa}, behavior monitoring of malicious patterns and system supervision \cite{AnBa}.

However, the increasing birth-rate of new or mutated malware samples has raised the need for efficient and elaborated malware detection techniques that can effectively detect new malware strains in reasonable amounts of time. The detection approaches are strongly connected to the features set provided through the previous stage of malware analysis, and are distinguished  to static and dynamic features, respectively. Static features may include, statistical analysis on n-grams or opcodes, properties of control flow graphs, while dynamic features are obtained the execution time of a program and concern its general behavior (i.e., interaction with the host-environment - O.S.), access events or any other interconnection patterns \cite{GrPaMaBaMc}.

Malware detection approaches are divided into two main categories, namely signature-based malware detection and behavior-based malware detection \cite{SoHo,DaDiViAuSt,AlSrDoMa,BeCiDiMaMe,HuChiShi,BaHaBaKiKr,TiBa,KeMo}. Next we briefly discuss these two methods and present some of the approaches deployed in each one.

\vspace*{0.05in}
\noindent \textbf{Signature-based Malware Detection.} Signature-based malware detection is the dominant technique deployed by antivirus software products due to its time efficacy that provides real-time protection against malicious threats \cite{DaFiGa}. A byte-level signature is a sequence (i.e., pattern) of bytes used to identify each newly discovered malware, using a scanning scheme of exact correlation and a repository of signatures in order to detect malicious software samples \cite{DaDiViAuSt}. A signature may represent a byte-code sequence, a binary assembly instruction, an imported Dynamic Link Library (DLL), or function and system calls. \cite{AlSrDoMa,BeCiDiMaMe}. Novel malware detection approaches using machine learning can be deployed through two methods, namely, assembly features and binary features \cite{SoHo}. However, signature-based detection techniques can easily be evaded through code obfuscation techniques that even the least modification on the code sequence would lead to a completely different byte-sequence \cite{DaDiViAuSt}. A major characteristic of signature-based malware detection is the exhibited precision so through object scanning utilizing efficient meta-heuristic algorithms as in the uniqueness signature creation. This characteristic regarding its precision may turn to a drawback, since such methods can not detect obfuscated or mutated (e.g., polymorphic) malware samples, as their signature does not match the stored one \cite{SoHo}.

\vspace*{0.05in}
\noindent \textbf{Behavior-based Malware Detection.} Another approach deployed for malware, gaining remarkable research interest during the last yeas is behavioral detection, or more formally, behavior-based malware detection \cite{JaDeFi}. Behavior-based malware detection mainly focuses on capturing the interaction (in terms of interconnection, relations or dependencies  between system-elements i.e., system-calls or API calls) between the executed software and the system (i.e., Operating System)\cite{BaReSo,ChJhKr,ChJhSeSoBr,FRJhChSaYa,BaMoKrKi,BeCiDiMaMe,BaCoHlKrKi,RiThCaPaLa,SaSgDiMa}. From an abstract machine learning aspect, the behavior-based systems are trained over a learning phase with behaviors exhibited during the execution of known malicious software samples, while in the monitoring phase the trained behavior-based system decides if an unknown software sample is malicious or not \cite{DaDiViAuSt}.
Behavior-based detection systems as expected require the execution of the software sample in order to extract dynamically (see, Dynamic Malware Analysis) the exhibited behaviors. In order for these dynamic systems to perform the mining of the specified behaviors they utilize software and hardware virtualization technologies, alongside with imitation conditions \cite{SoHo}, providing the test sample with an environment as close to reality in order to evade the sandbox-detection mechanisms deployed occasionally by malicious software samples, and letting them exhibit their intentions. Despite the fact that such techniques deploy quite elaborate algorithms on their implementations, the incident that malware families tend to evolve in order to avoid detection \cite{AlSrDoMa}, results to the need of the development of more elastic and mutation resilient techniques like the one we propose in this work.

\vspace*{0.1in}
\subsection{Classifying Malware Samples}

\vspace*{0.05in}
\noindent Malware authors, in order to avoid traditional detection methods, produce new (mostly mutated) malware samples rapidly, utilizing existing ones in order for the new strains to preserve the functionality inherited from their ancestors. As referred in \cite{BeCiDiMaMe} mutated malware samples are generated from existing ones utilizing automated techniques \cite{TiBaVe,YoYi} or integrated tools, generating new samples from libraries and code parts from code exchange networks.

\vspace*{0.05in}
\noindent \textbf{Malware Classification.} Through the literature, the term {\it malware classification} has been confused several times with {\it malware detection}. Distinguishing precisely these two procedures, it can be stated that malware detection is a {\it binary classification}, where a a set of unknown samples is classified against a collection of malware and goodware samples, while malware classification  is a {\it multinomial classification} on whether an already detected malware sample belongs to a particular family or type \cite{GrShFr}. As described in \cite{KoErWeZaEc}, malicious software samples that belong to the same malware family tend to exhibit similar behavioral and structural profiles. Additionally, malware classification augments the analysis of new, or mutated, malicious samples where their signatures have not been constructed yet \cite{PaReMuSu}.

\vspace*{0.05in}
\noindent \textbf{Malware Phylogeny.} Another field of malware analysis applied in malware classification is {\it malware phylogeny}\cite{MoChFi}, which aims on inferring evolutionary relationships between instances of families. The major profit from creating a phylogeny model is the fact that newly developed elaborated detection systems that deploying such techniques can detect that a sample that has not been previously seen can be related to a malware family, when analyzed along an evolution path \cite{LiDaLiWa}. Throughout this process the main target is to reveal similarities and relations among a set of specific malware samples coexist and are exhibited by all the members of the set (i.e., malware family) \cite{SuChHsCh}, distinguishing its type or family. Such approaches can be utilized to identify evolution trends in over a set of malware samples \cite{BeCiDiMaMe}, constituting hence valuable tool for more generalized signatures or, in general, more elaborated detection-techniques. The models applied on phylogeny, using mostly phylogenetic networks, model evolutionary relations among malware families, describing temporal ordering among samples, defining ancestor-descendent relations, as also relationships between families, augmenting hence malware classification \cite{} and unveiling evolutionary trends \cite{LiWaDaWa}.

\vspace*{0.2in}
\section{Model Components}
\label{sec:Components}
\vspace*{0.1in}

\noindent In this section, we firstly discuss the properties of our initial behavioral-graph representation i.e., the System-call Dependency Graph (or, for short, ScDG) and the proposed structural components of our model, namely, the Group Relation Graph (or, for short, GrG) and Coverage Graph (or, for short, CvG). In order to invoke the temporal evolution of these primordial graphs, we propose and present the derivative graphs the depict the temporal evolution of GrG and CvG, namely the Group Relation Temporal Graph (or, for short, GrTG) and Coverage Temporal Graph (or, for short, CvTG), respectively. More precisely, we show how the system-calls invoked through the execution of a program consisting its ScDG, are merged into groups of similar functionality constructing a directed edge-weighted graph called GrG that its vertex set refers to a system-call group, while its edge set contains the interconnection between the system-calls of these groups, and how we construct its corresponding component, i.e., the CvG, which is a vertex-weighted undirected graph. Given such graph representations, we present the construction of the key components of our proposed detection and classification model, i.e., their corresponding derivative graphs GrTG and CvTG, depicting their temporal evolution.

\vspace*{0.1in}
\subsection{System-call Dependency Graphs}

\begin{figure}[t!]
\hrule \smallskip
    \begin{minipage}[r]{1.8in}
        \vspace*{0.3in}
        \includegraphics[scale=0.45]{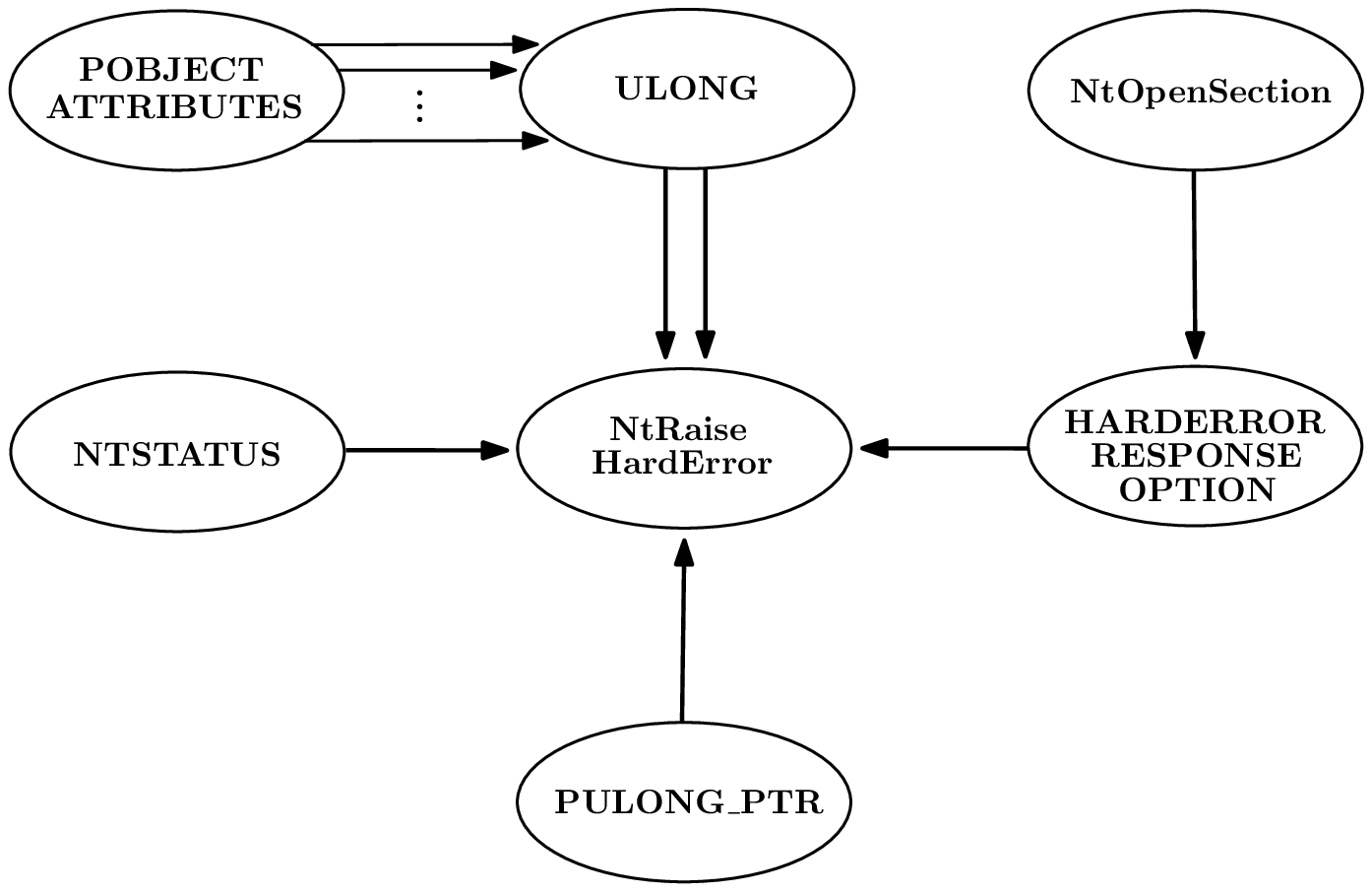}
    \end{minipage}
    \hspace*{2in}
    \begin{minipage}[r]{1.8in}
       \vspace*{0.3in}
       \includegraphics[scale=0.42]{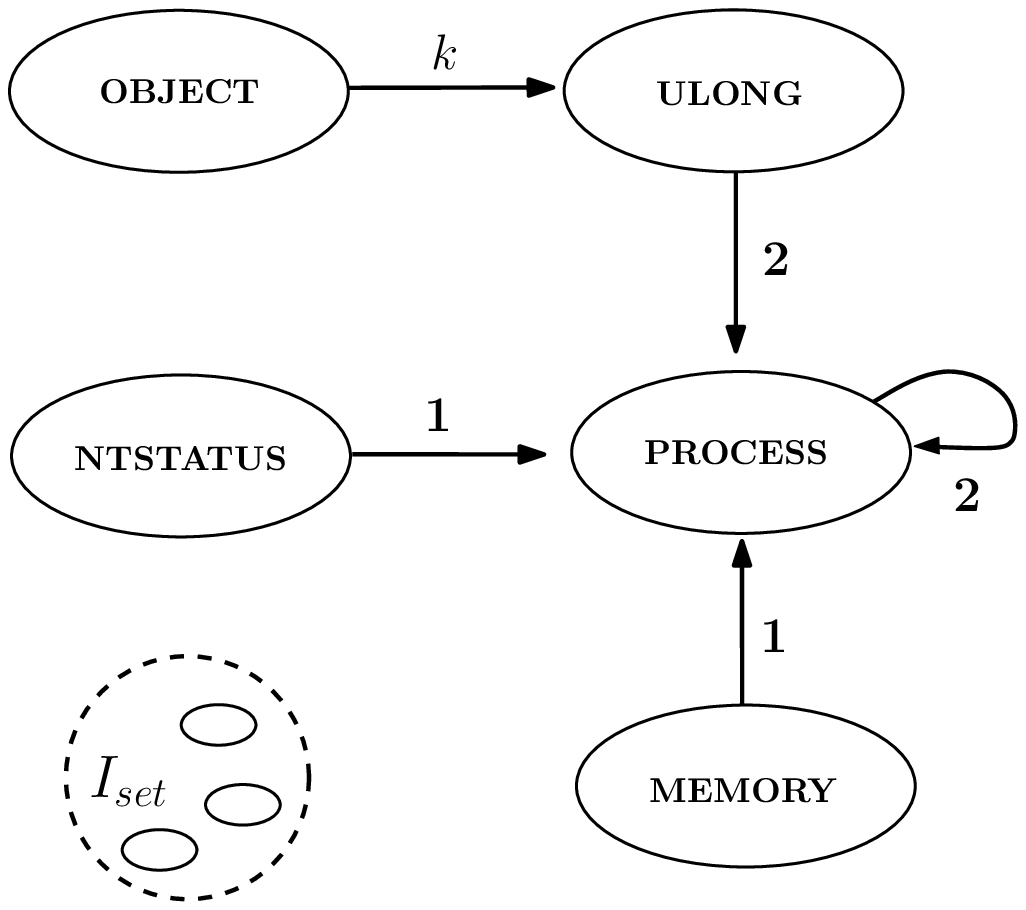}
    \end{minipage}\\ \\
    {\hspace*{1.1 in} {\small (a)}} \hspace*{3.1in} {\small (b)}\\
    \hrule\smallskip
    \caption{\small{(a) The System-call Dependency Graph of a program; (b) The corresponding Group Relation Graph of a program.}}
\label{fig:fig1}
\end{figure}

\vspace*{0.05in}
\noindent The system-calls invoked during the execution of a program can be traced through taint analysis, and hereafter the behavior of a program can be represented with a directed acyclic graph (dag), the so called System-call Dependency Graph see, Figure~\ref{fig:fig1}(a). The vertex set of a ScDG is consisted by all the system-calls invoked during the execution of a program and its edge set represents the communication between system-calls as described in \cite{NiPo,BaReSo,FRJhChSaYa}.

Recalling that the suspicious sample needs to be executed in a contained environment (i.e., a virtual machine), where during its execution time, dynamic taint analysis is performed in order to capture system-call traces, next we illustrate a simple example that includes the system-call traces obtained, constructing the ScDG of a program. In Figure~\ref{fig:fig1}(a), it is easy to see that the vertex set of this graph is consisted from the system-calls invoked during the execution of the sample and its edge set is consisted by their in between data-flow dependencies, constructing a directed acyclic graph (dag).

\vspace*{0.1in}
\subsection{Group Relation Graphs}

\vspace*{0.05in}

\noindent Given a graph representation of malware-behavior such a ScDG, a more abstract graph representation of a program's behavior can be constructed based on the fact that system-calls of similar functionality can be classified into the same group (see Table\ref{Tab1}). The produced graph representation is a directed weighted graph called Group Relation Graph; see, Figure~\ref{fig:fig1}(b). The whole procedure for constructing the GrG graph from a given ScDG for a program is described in details in \cite{NiPo,NiPo2}.

\begin{table}[t!]
\centering
\smallskip\smallskip
\begin{tabular}{|l|c|l|c|}
\hline
\textbf{\ Group Name} & \textbf{Size} & \textbf{\ Group Name} & \textbf{Size} \\ \hline \hline
\ ACCESS\_MASK \phantom{XX}	&	1	&	\ PHANDLE		&	1	\\	\hline
\ Atom	&	5	& \ PLARGE\_INTEGER		&	1	\\	\hline
\ BOOLEAN	&	1	& \ Process		&	49	\\	\hline
\ Debug	&	17	& \ PULARGE\_INTEGER	\phantom{XX}	&	1	\\	\hline
\ Device	&	31	& \ PULONG		&	1	\\	\hline
\ Environment	&	12	& \ PUNICODE\_STRING		&	1	\\	\hline
\ File	&	44	& \ PVOID\_SIZEAFTER		&	1	\\	\hline
\ HANDLE	&	1	& \ PWSTR		&	1	\\	\hline
\ Job	&	9	& \ Registry		&	40	\\	\hline
\ LONG	&	1	& \ Security		&	36	\\	\hline
\ LPC	&	47	& \ Synchronization		&	38	\\	\hline
\ Memory	&	25	& \ Time		&	5	\\	\hline
\ NTSTATUS	&	1	& \ Transaction		&	49	\\	\hline
\ Object	&	19	& \ ULONG		&	1	\\	\hline
\ Other	&	36	& \ WOW64		&	19	\\	\hline
\end{tabular}
\vspace{0.1in}
\caption{The 30 system-call groups - Total Groups.}
\vspace*{-0.2in}
\label{Tab1}
\end{table}
\vspace*{0.2in}

As described in \cite{NiPo}, having the grouping of system-calls and a system-call dependency graph ScDG, the GrG graph $\widehat{G}$ is a directed edge-weighted graph on $n$ vertices $v_1$, $v_2$, $\ldots$, $v_{n}$ constructed as follows:

\begin{itemize}
\item[(i)] for every pair $\{v_i, v_j\} \in V(\widehat{G})$, a directed edge~$(v_i, v_j)$ is added in $E(\widehat{G})$ if the two system-calls communicating with each other, let $(S_p,S_q)$, is an edge in $E(ScDG)$ and, $S_p$ belongs to the $i$-th system-call group and $S_q$ belongs to the  $j$-th system-call group;
\vspace{-0.0in}
\item[(ii)] for each directed edge $(v_i, v_j) \in E(\widehat{G})$, a weight $w(v_i, v_j) \in \Re$ is assigned on it if there are $w(v_i, v_j)$ invocations from a system-call in the $i$-th group to a system-call in the $j$-th group.
\end{itemize}

\noindent Having defined the GrG graph $\widehat{G}$, we also define the underlying vertex-weighted graph $\widetilde{G}$ of the  graph $\widehat{G}$ having vertex-weights $w(v_i)=\sum_{v_j \in Adj(v_i)} w(v_i,v_j)$, for every $v_i \in V(\widetilde{G})$.

\begin{figure}[t!]
\hrule \smallskip
    \begin{minipage}[r]{1.8in}
        \vspace*{0.3in}
        \includegraphics[scale=0.4]{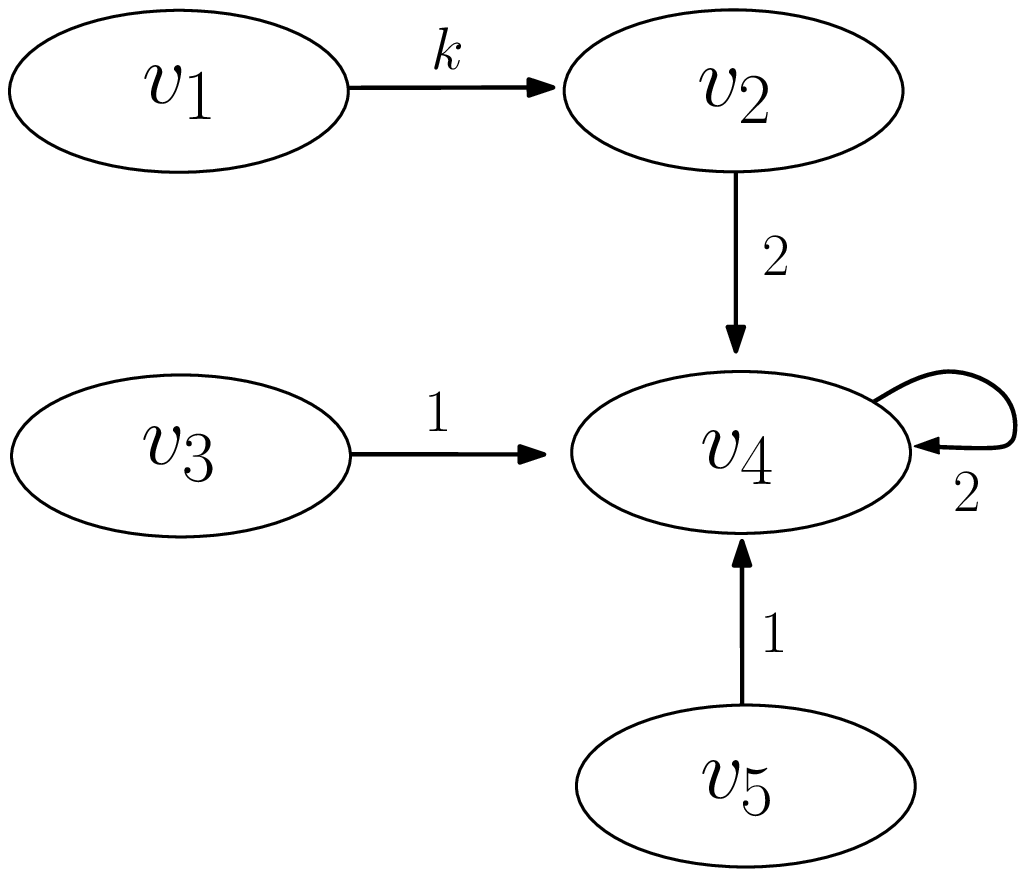}
    \end{minipage}
    \hspace{0.2in}
    \begin{minipage}[r]{1.8in}
        \vspace*{0.3in}
        \includegraphics[scale=0.4]{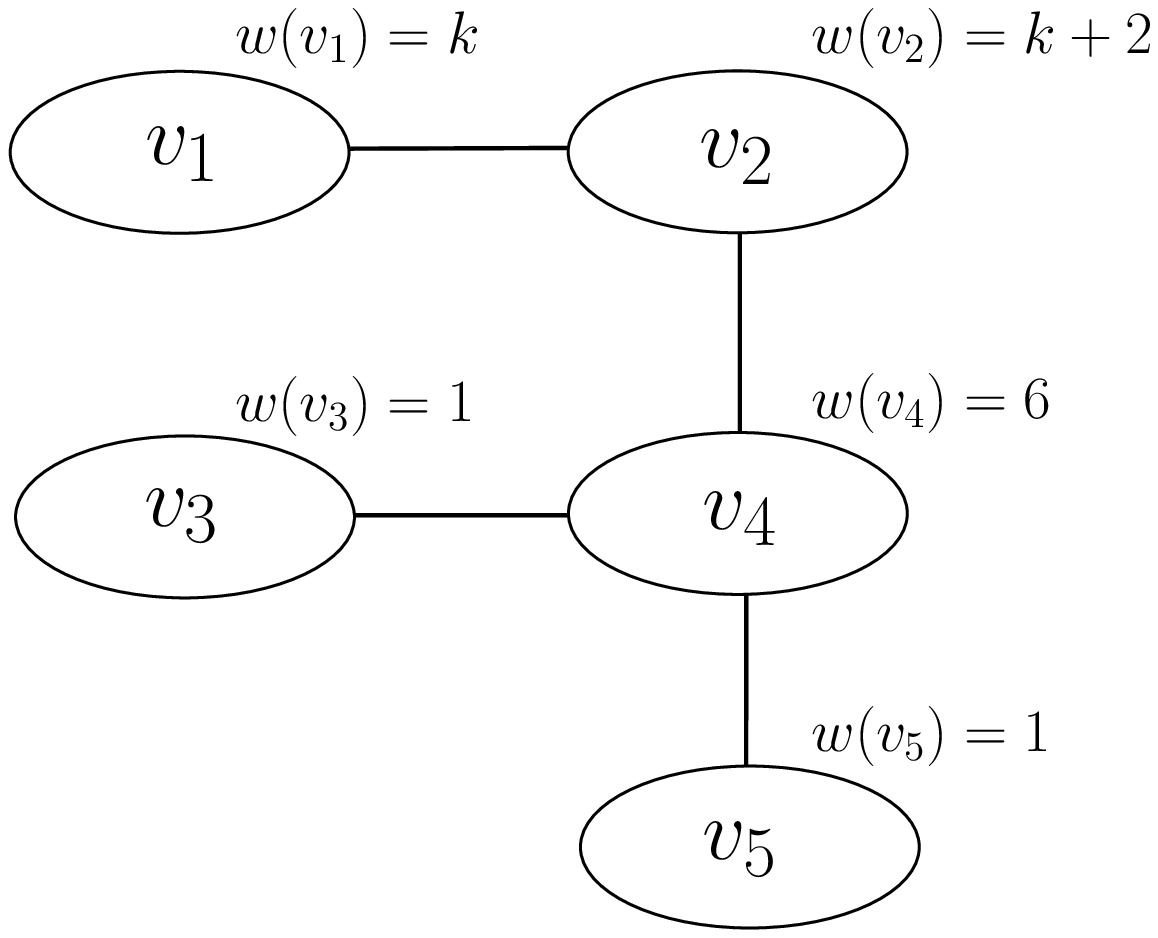}
    \end{minipage}
    \hspace{0.3in}
    \begin{minipage}[r]{1.8in}
       \vspace*{0.4in}
       \includegraphics[scale=0.4]{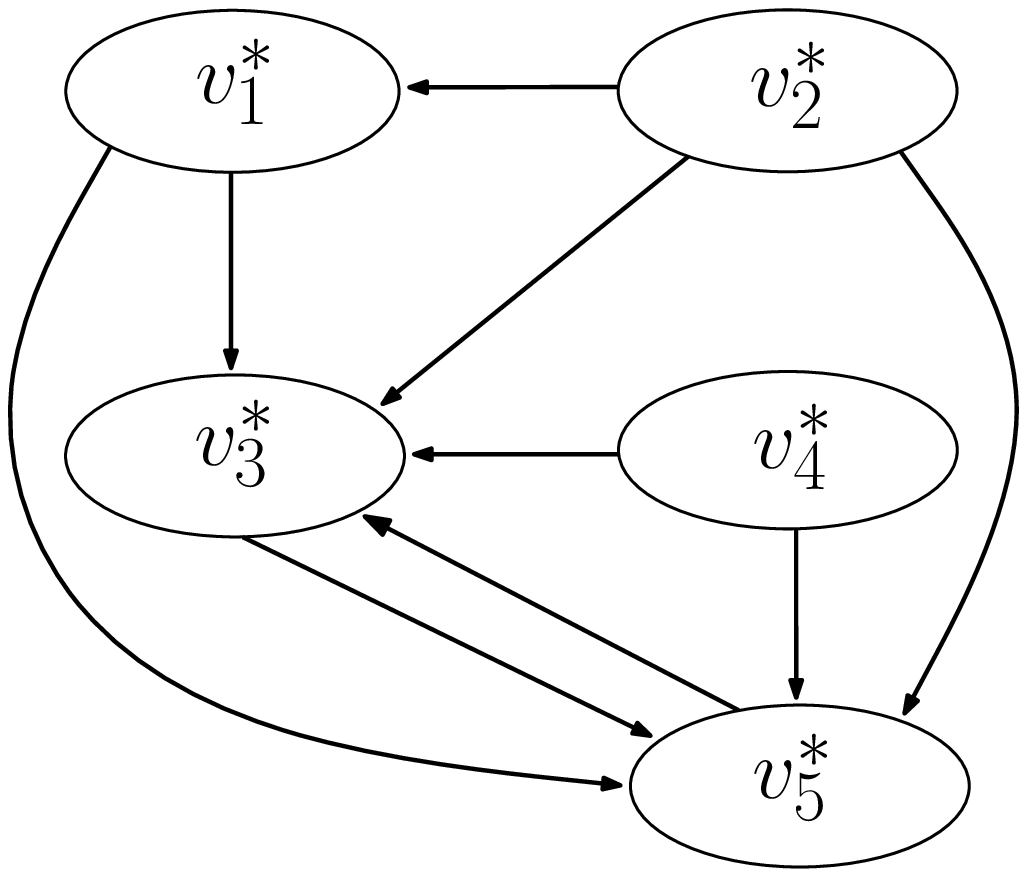}
    \end{minipage}\\ \\
    {\hspace*{0.6 in} {\small (a)}} \hspace*{1.8 in} {\small (b)} \hspace*{2.0 in} {\small (c)}\\
    \hrule\smallskip
    \caption{\small{(a) A GrG graph $\widehat{G}$; (b) Its underlying vertex-weighted graph $\widetilde{G}$; (c) The CvG graph $G^*$ produced by the graph $\widetilde{G}$ through its vertex domination relations.}}
\label{fig:fig2}
\end{figure}

\vspace*{0.1in}
\subsection{Coverage Graphs}

\vspace*{0.05in}
\noindent Another component of our detection model is the Coverage Graph (or, for short, CvG) \cite{MpNiPo}. As we mentioned above, a GrG graph $\widehat{G}$ is an edge-weighted directed graph which, in our approach, we transform it to its underlying vertex-weighted undirected graph $\widetilde{G}$. We first define domination relations on the vertices of the graph $\widetilde{G}$ and then utilizing these relations we construct the Coverage Graph of the GrG graph $\widehat{G}$, denoted by $G^*$.

We also present a 2D-representation of the underlying vertex-weighted graph $\widetilde{G}$ of the graph $\widehat{G}$ utilizing the degrees and the vertex-weights of its vertices and show a deferent way to compute the domination relations on the graph $\widetilde{G}$.

\begin{definition}
Let $\widetilde{G}$ be the underlying vertex-weighted graph of a GrG graph $\widehat{G}$ and let $v_i$, $v_j \in V(\widetilde{G})$. We say that $v_i$ {\it dominates} $v_j$, denoted by $v_i \xrightarrow[\text{}]{dom} v_j$, if $deg(v_i) \geq deg(v_j)$ and $w(v_i) \geq w(v_j)$, where $deg(v)$ and $w(v)$ denote the degree and the weight of the vertex $v \in V(\widetilde{G})$, respectively.
\end{definition}

\noindent  The domination set $D_i$ of a vertex $v_i \in V(\widetilde{G})$ is the set of all the vertices $v_j: v_i \xrightarrow[\text{}]{dom} v_j$. If $v_i \xrightarrow[\text{}]{dom} v_j$ we say that vertices $v_i$ and $v_j$ are in a domination relation.

\begin{definition}
Let $\widetilde{G}$ be the underlying vertex-weighted graph of a GrG graph $\widehat{G}$ with vertices $V(G)=\lbrace v_1, v_2, \cdots, v_n \rbrace$. The Coverage Graph (CvG) of the GrG graph $\widehat{G}$, denoted also $G^*$, is a directed graph defined as follows:

\begin{itemize}
\item[(i)] $V(G^*)=\lbrace v^*_1, v^*_2, \cdots, v^*_n \rbrace$ and
$\lbrace v^*_1, v^*_2, \cdots, v^*_n \rbrace \leftrightarrow \lbrace v_1, v_2, \cdots, v_n \rbrace $;
\vspace{-0.0in}
\item[(ii)] $v^*_i v^*_j \in E(G^*)$ if $v_i \xrightarrow[\text{}]{dom} v_j$,
 where $v_i$ and $v_j$ correspond to $v^*_i$ and $v^*_j$, respectively.
\end{itemize}

\end{definition}

\noindent In Figure~\ref{fig:fig2}(a) we show a GrG graph $\widehat{G}$ which is isomorphic to the GrG graph $\widehat{G} \backslash I_{set}$ of Figure~\ref{fig:fig1}(b), in Figure~\ref{fig:fig2}(b) we depict its underlying vertex-weighted graph $\widetilde{G}$ with $w(v_i)=\sum_{v_j \in Adj(v_i)} w(v_i,v_j)$, $\forall v_i \in V(\widetilde{G})$, where $w(v_i,v_j)$ is the weight of the edge $(v_i,v_j) \in E(\widehat{G})$, while in Figure~\ref{fig:fig2}(c) we show the Coverage Graph $G^*$ constructed from the graph $\widetilde{G}$ by utilizing its vertex domination relations. Note that the vertices of each graph in Figure~\ref{fig:fig2} correspond the System-call groups of Figure~\ref{fig:fig1}(b).

\begin{figure}[t!]
\hrule
    \begin{minipage}[r]{1.8in}
        \vspace*{0.4in}
        \includegraphics[scale=0.40]{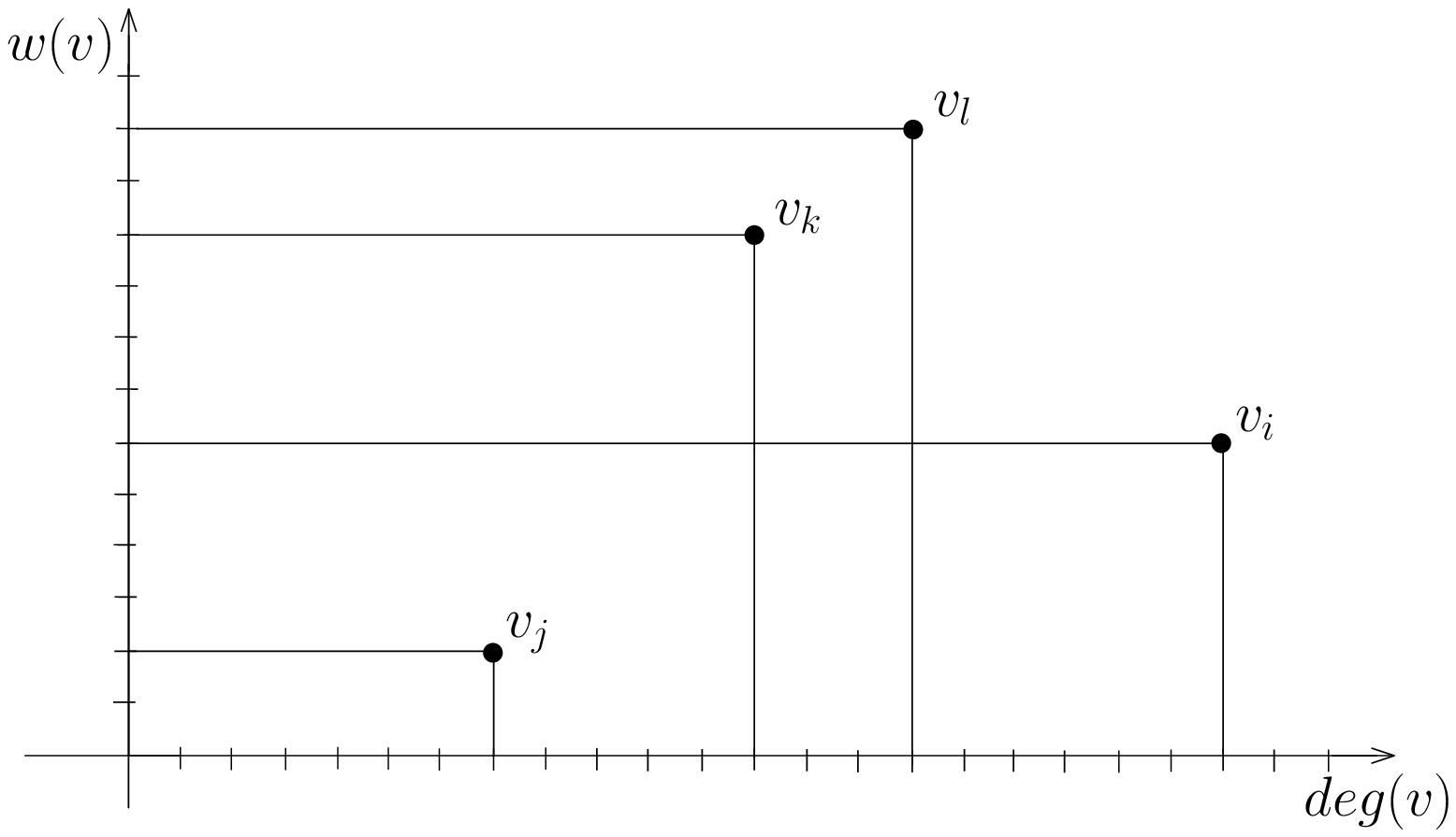}
    \end{minipage}
    \hspace*{1.5in}
    \begin{minipage}[r]{1.8in}
       \vspace*{0.5in}
       \includegraphics[scale=0.38]{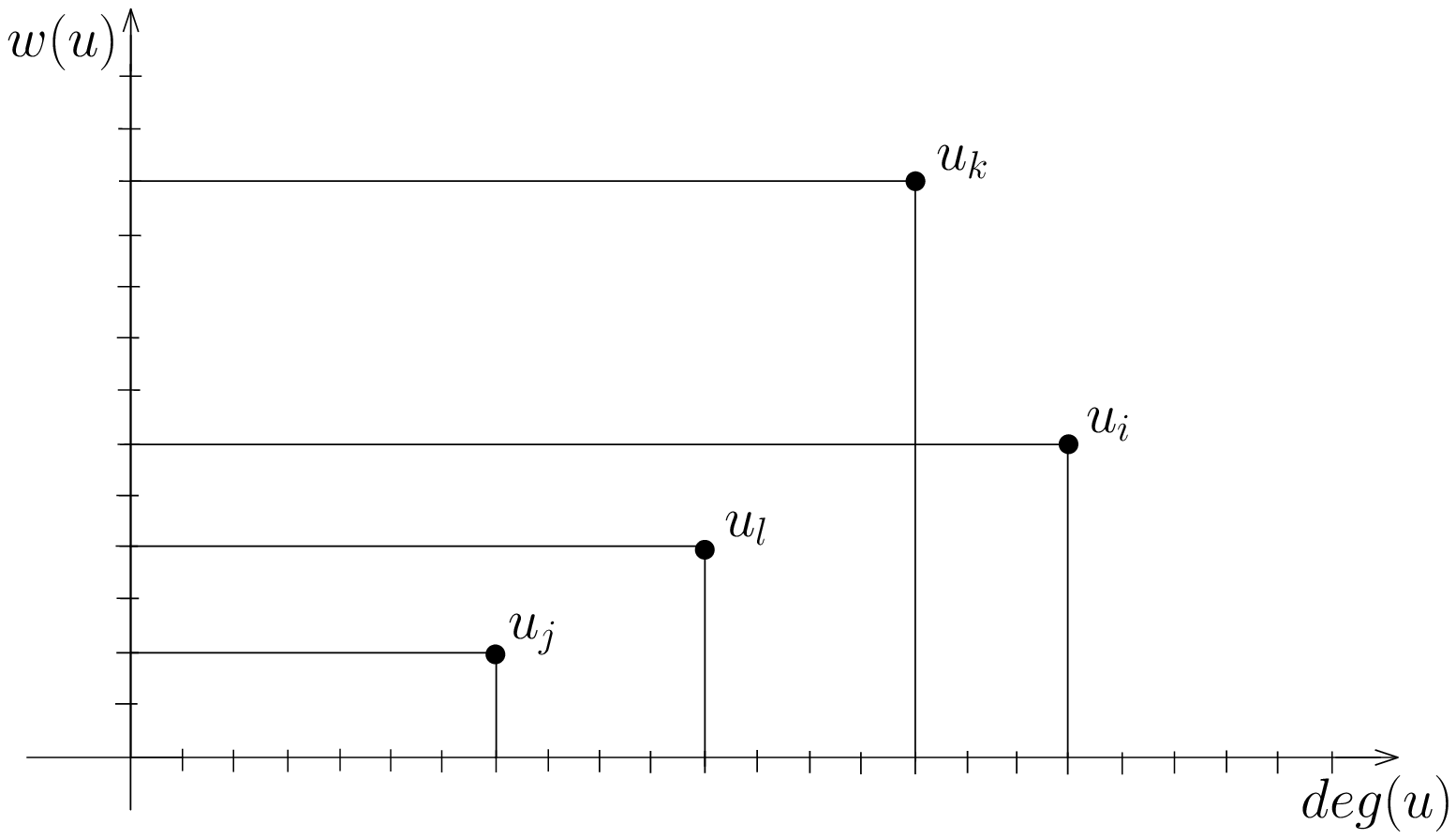}
    \end{minipage}\\ \\
    {\hspace*{1.1 in} {\small (a)}} \hspace*{3.1in} {\small (b)}\\
    \hrule\smallskip
    \caption{\small{(a) The 2D-representation of the graph $\widetilde{G}$ of a GrG graph $\widehat{G}$ produced by the ScDG of an unknown software; (b) The 2D-representation for the case where the ScDG graph is a known malicious software.}}
\label{fig:fig3}
\end{figure}

\noindent In Figure~\ref{fig:fig3} we show a 2D-representation of the underlying vertex-weighted graph $\widetilde{G}$ of a GrG graph $\widehat{G}$. The 2D-representation in Figure~\ref{fig:fig3}(a) depicts the domination relations on the vertices of the test sample's GrG, while the one in Figure~\ref{fig:fig3}(b) depicts the domination relations on the vertices of the malware sample's GrG. As we can observe, according to the definition of the domination relation, in Figure~\ref{fig:fig3}(a) the domination sets of the vertices $v_i$, $v_j$, $v_k$, and $v_l$ are $D_i=\lbrace v_j \rbrace$, $D_j=\emptyset$, $D_k=\lbrace v_j \rbrace$, and $D_l=\lbrace v_k,v_j \rbrace$, respectively, while in Figure~\ref{fig:fig3}(b) the vertex domination sets of the corresponding vertices $u_i$, $u_j$, $u_k$, and $u_l$ are $D'_i=\lbrace u_l,u_j \rbrace$, $D'_j=\emptyset$, $D'_k=\lbrace u_l,u_j \rbrace$, and $D'_l=\lbrace u_j \rbrace$, respectively.

\vspace*{0.1in}
\subsection{Temporal Graphs}

\vspace*{0.05in}

\noindent Throughout the development of our research, we have noticed that, to the best of our knowledge, there does not exist any approach on the literature that references or leverages the factor of the temporal evolution of a graph. Similarly to {\it philogeny} that examines the temporal evolution of malware families, the key component of our proposed detection and classification model, leverage the temporal evolution of graphs (i.e., GrG and CvG graphs) in order to depict the structural modifications performed on the graph and that could distinguish either a malware sample or to a further extent a malware family.

In our model, we define two types of graphs that depict the temporal evolution of our initial graph structures (i.e., Group Relation Graphs and Coverage Graphs), namely Group Relation Temporal Graphs or, for short, GrTG and Coverage Temporal Graphs or, for short, CvTG, respectively. In order to implement such graph structures we approach this modeling by creating instances of the initial GrG and CvG graphs during their construction. As we mentioned above, GrG graphs are constructed by the sum of the system-calls invoked interconnecting pairs of system call groups, and correspondingly CvG are constructed by they respective dominating relation (i.e., by their supremacy regarding degree and weight) between the system-call groups. Hence, since we are given the system-call dependencies in a series that depicts the time correlation among (i.e., an edge sequence of the System-call Dependency Graph that shows the system-call invocations during execution time), such constructions can be obtained by creating an instance of the produced graphs (i.e., GrG, CvG) at specific steps.

Formalizing our previous claim, we can define that for a set of time-slots, let $t_1, t_2, \ldots, t_n$ we can construct $n$ instances of graphs GrG and CvG and denote them as $T_1(\widehat{G}), T_2(\widehat{G}), \ldots, T_n(\widehat{G})$ and $T_1(G^*), T_2(G^*), \ldots, T_n(G^*)$, respectively, that depict the structure in terms of edges, vertex-degrees and vertex-weights of the corresponding graphs at specific time slots. Through this approach we can maintain information about the temporal evolution of the graph thorough its construction procedure,and further leverage such information in order to perform more elaborated graph similarity techniques.

\begin{figure}[t!]
\hrule \smallskip
    \begin{minipage}[r]{1.8in}
        \vspace*{0.3in}
        \includegraphics[scale=0.7]{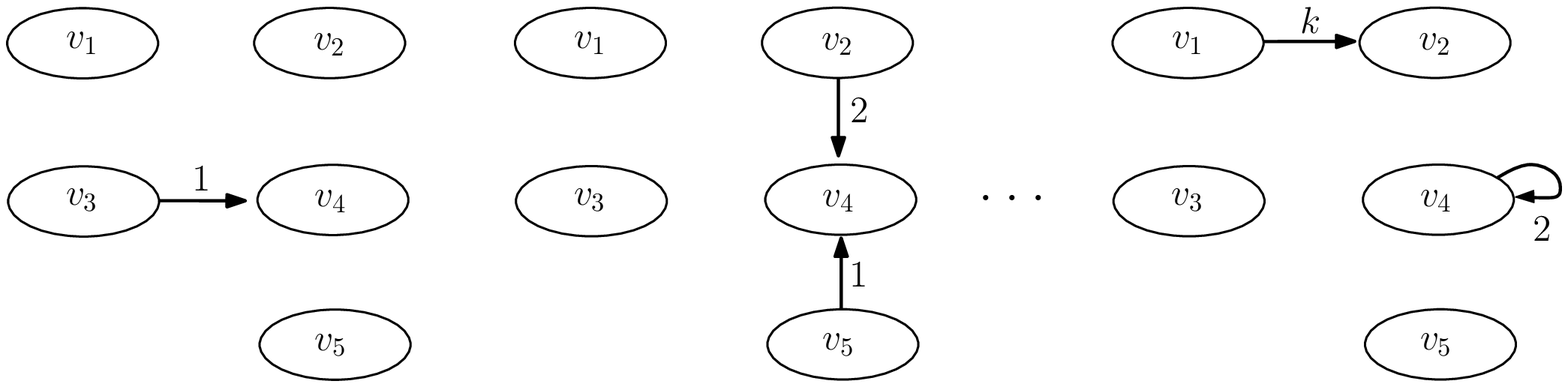}
    \end{minipage}
    \\ \\
    {\hspace*{0.6 in} {\small (a)}} \hspace*{1.6 in} {\small (b)} \hspace*{2.0 in} {\small (c)}\\
    \hrule\smallskip
    \caption{\small{\small{The temporal evolution of a GrG graph $\widehat{G}$ represented by its Discrete Modification Temporal Graph $T^{f}(\widehat{G})$ over $n$ {\tt epochs}:  (a) $T^{f}_{1}(\widehat{G})$, (b) $T^{f}_{2}(\widehat{G})$ and (c) $T^{f}_{n}(\widehat{G})$.}}}
\label{fig:fig4a}
\bigskip\medskip\smallskip
\hrule \smallskip
    \begin{minipage}[r]{1.8in}
        \vspace*{0.3in}
        \includegraphics[scale=0.7]{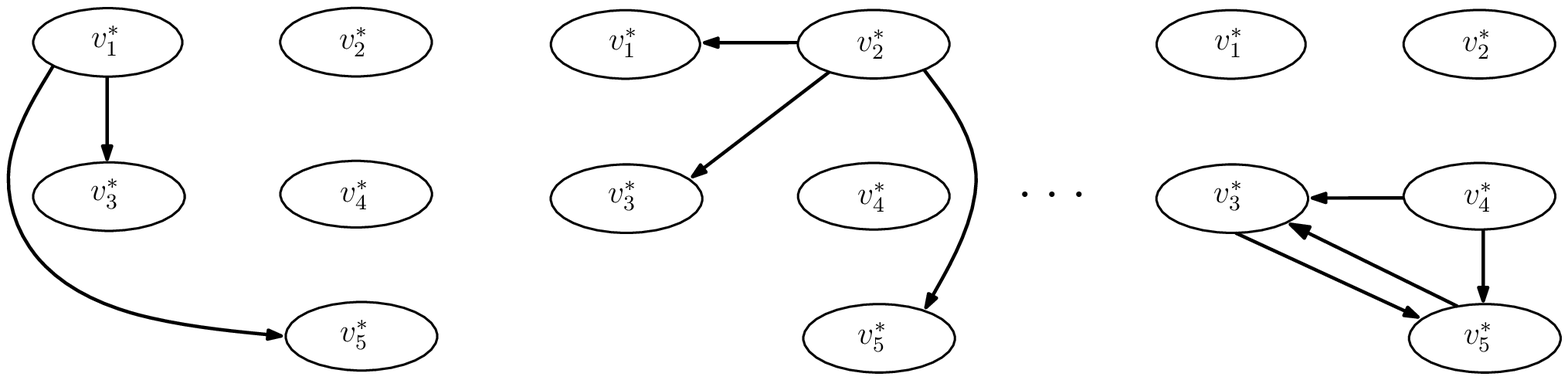}
    \end{minipage}
    \\ \\
    {\hspace*{0.6 in} {\small (a)}} \hspace*{1.6 in} {\small (b)} \hspace*{2.0 in} {\small (c)}\\
    \hrule\smallskip
    \caption{\small{The temporal evolution of a CvG graph $G^*$ represented by its Discrete Modification Temporal Graph $T^{f}(G^*)$ over $n$ {\tt epochs}:  (a) $T^{f}_{1}(G^*)$, (b) $T^{f}_{2}(G^*)$ and (c) $T^{f}_{n}(G^*)$.}}
\label{fig:fig5a}
\end{figure}

\vspace*{0.05in}
\noindent \textbf{Partitioning Time.} The factor of time actually does not represent the actual quantum of run-time, but each time-quantum corresponds to one system-call dependency or, equivalently, relation between two System-call Groups (i.e., edge of the Group Relation Graph). Hence, the total time-line depicts the slots or time-partitions from the appearance of the first to the last group relation.

Additionally, in our model, we define as {\tt epochs} the set of time-partitions, i.e., $t_1, t_2, \ldots, t_n$, and an {\tt epoch}, let $t_i$, contains the structural modifications (i.e., edges added on the corresponding GrG or CvG graph) from the begin to the end of the $i^{th}$ {\tt epoch}, where $1<i<n, \forall n \in D$, and $D=\lbrace \forall n \in N : |E(G)|\mod n=0 \rbrace$.

As we described throughout the paper, the conceptual substance of Temporal Graphs is to depict the structural evolution of the GrG and CvG graphs through the time. However, the structural modification on the instances of the graph over the time can be described either discretely as addition of edges over the exact previous graph instance, or cumulatively as successive additions of edges performed on all the previous graph instances. Next, we discuss the construction of the corresponding Temporal Graphs according to the two approaches.

\vspace*{0.05in}
\noindent \textbf{Discrete Modification Temporal Graphs.} In the first approach of our proposed scheme, the construction of the Temporal Graph, that represents the evolution of GrG or CvG graphs during time, constructs the induced subgraph of GrG and CvG, respectively, including only the edges that where added on a specific {\tt epoch}. So, let the {\tt epoch} $t_i$ we construct the Temporal Graphs $GrTG_i$, $CvTG_i$ of the graphs GrG and CvG, denoting them with $T^{f}_{i}(\widehat{G})$, $T^{f}_{i}(G^*)$, respectively, where $f$ denotes the cardinality of edges added on this {\tt epoch}. In Figure~\ref{fig:fig4a} and Figure~\ref{fig:fig5a}, we depict the discrete structural modification (i.e., temporal evolution) of graphs GrG and CvG over the construction of their corresponding Temporal Graphs $T^f(\widehat{G})$ and $T^f(G^*)$ during $n$ {\tt epochs}.

\begin{figure}[t!]
\hrule \smallskip
    \begin{minipage}[r]{1.8in}
        \vspace*{0.3in}
        \includegraphics[scale=0.7]{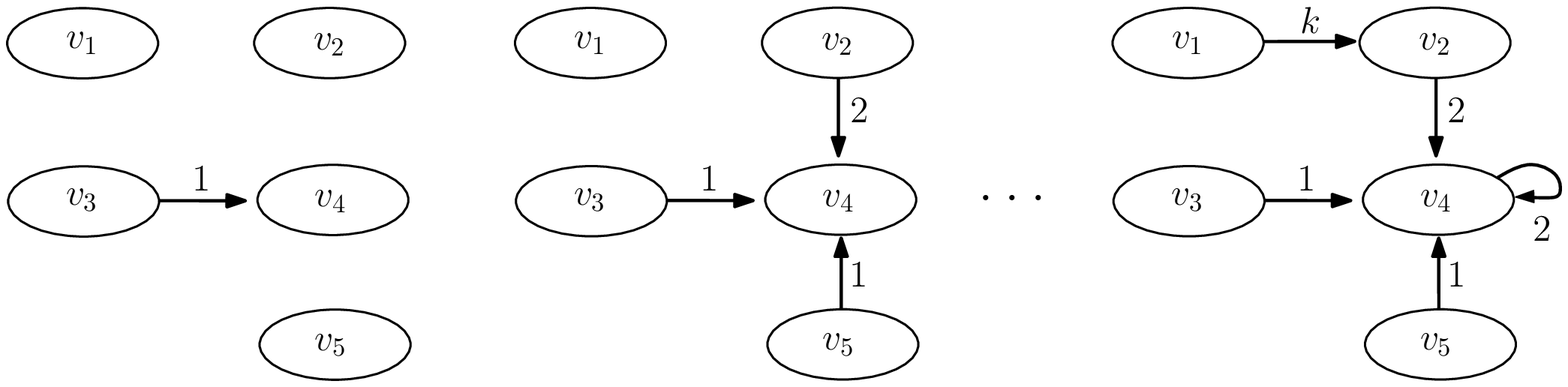}
    \end{minipage}
    \\ \\
    {\hspace*{0.6 in} {\small (a)}} \hspace*{1.6 in} {\small (b)} \hspace*{2.0 in} {\small (c)}\\
    \hrule\smallskip
    \caption{\small{The temporal evolution of a GrG graph $\widehat{G}$ represented by its Cumulative Modification Temporal Graph $T^{F}(\widehat{G})$ over $n$ {\tt epochs}:  (a) $T^{F}_{1}(\widehat{G})$, (b) $T^{F}_{2}(\widehat{G})$ and (c) $T^{F}_{n}(\widehat{G})$.}}
\label{fig:fig4b}
\bigskip\medskip\smallskip
\hrule \smallskip
    \begin{minipage}[r]{1.8in}
        \vspace*{0.3in}
        \includegraphics[scale=0.7]{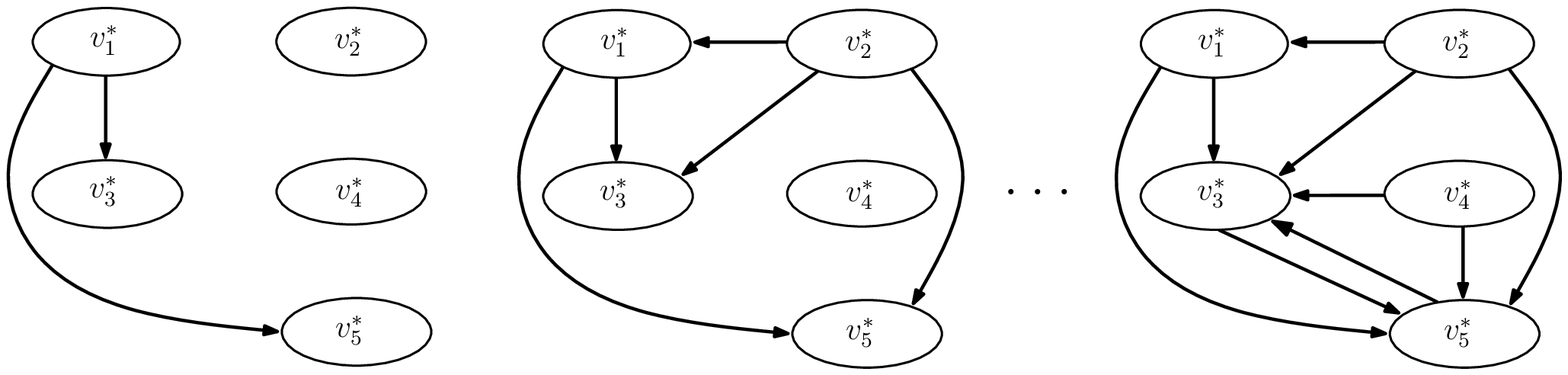}
    \end{minipage}
    \\ \\
    {\hspace*{0.6 in} {\small (a)}} \hspace*{1.6 in} {\small (b)} \hspace*{2.0 in} {\small (c)}\\
    \hrule\smallskip
    \caption{\small{The temporal evolution of a CvG graph $G^*$ represented by its Cumulative Modification Temporal Graph $T^{F}(G^*)$ over $n$ {\tt epochs}:  (a) $T^{F}_{1}(G^*)$, (b) $T^{F}_{2}(G^*)$ and (c) $T^{F}_{n}(G^*)$.}}
\label{fig:fig5b}
\end{figure}

\vspace*{0.05in}
\noindent \textbf{Cumulative Modification Temporal Graphs.} In this type of Temporal Graphs, the evolution of the graph is represented as an additive procedure, since once an edge has been created at a given time, let $i$,  on the graph between two system-call groups on the GrG graph, or a domination relation has been resulted on the CvG graph, it will remain permanent on the ancestor Temporal Graphs (i.e., if $\lbrace u,v \rbrace \in E(T_i(\widehat{G})) \rightarrow \lbrace u,v \rbrace \in E(T_j(\widehat{G})$ and if $\lbrace u,v \rbrace \in E(T_i(G^*)) \rightarrow \lbrace u,v \rbrace \in E(T_j(G^*))$, $\forall i<j<n$), since it consists a predecessor of the following temporal graphs. In the second approach of our proposed scheme, the construction of the Temporal Graph, that represents the evolution of GrG or CvG graphs during time, actually extends the graphs GrG and CvG, respectively, during time, by adding on them the edges that where added on a specific {\tt epoch}. So, let the {\tt epoch} $t_i$ we construct the Temporal Graphs $T(\widehat{G})$, $T(G^*)$ of the graphs GrG and CvG, denoting them with $T^{F}_{i}(\widehat{G})$, $T^{F}_{i}(G^*)$, respectively, where $F$ denotes the cardinality of edges added from {\tt epoch} $t_1$ until {\tt epoch} $t_i$. In Figure~\ref{fig:fig4b} and Figure~\ref{fig:fig5b}, we depict the cumulative structural modification (i.e., temporal evolution) of graphs GrG and CvG over the construction of their corresponding Temporal Graphs $T^F(\widehat{G})$ and $T^F(G^*)$ during $n$ {\tt epochs}.

\vspace*{0.2in}
\section{System Architecture}
\label{sec:Architecture}
\vspace*{0.1in}

\noindent In this section we present the key components of our detection and classification model, and describe the key insights that constitute the basis of the corresponding procedures. Discussing the design principles that rule the deployment of our model's components, we present an overview of our detection and classification techniques.

\vspace*{0.1in}
\subsection{Design Principles}

\begin{figure}[t!]
    \centering
    \hrule\medskip\medskip
    \includegraphics[scale=0.5]{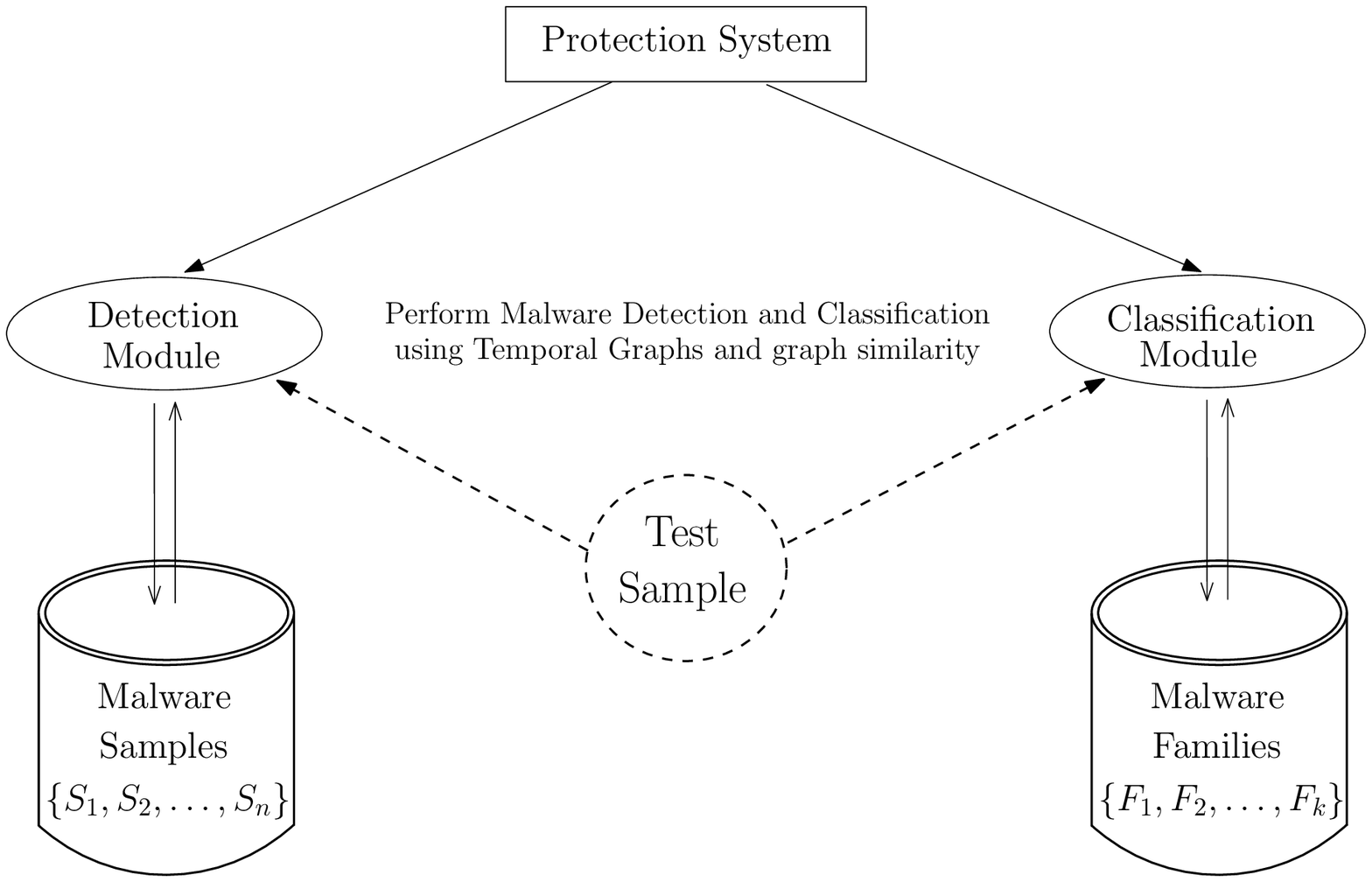}
    \vspace{0.25cm}
    \medskip\hrule
    \vspace{0.1cm}
    \caption{Architecture of the proposed system for detection and classification of malicious software.}
    \vspace{0.1cm}
  \label{fig:fig6}
\end{figure}

\vspace*{0.05in}
\noindent Malware detection and classification are two interconnected procedures. In malware detection the main target is to determine whether a given program is malicious or benign according to something that is known to be malicious, where malware classification is the following procedure and its intent is to determine the malware family to which the sample, that has been detected as malicious, belongs to. It easily follows that, an {\it a priori} knowledge of characteristics of known maliciousness has to be stored in a knowledge database, as also that in order to compare two subjects a similarity measure among them is needed. Moreover, the proposed theoretical approach specifies the form of the subjects, where graph-based models interact with similarity metrics that measure the qualitative characteristics that represent evolutionary commonalities between temporal graphs.
Next, we present the architectural considerations, the design principles, the functionality, and the corresponding deployment of the key components of our proposed detection and classification model (see, Figure~\ref{fig:fig6}).

\vspace*{0.05in}
\noindent \textbf{Knowledge Database.} The knowledge database is consisted by a set of known malicious samples that have been classified to malware families according to their functional, structural and mostly behavioral commonalities. More precisely, various anti-virus vendors have classified these samples to families based on their own heuristic rules concerning shared behavioral patterns and functionally similar execution profiles. Regarding the detection process, the knowledge database, except the known malicious samples, also includes benign samples in order to measure the false positive rates (i.e., benign samples that have been detected as malicious) evaluating the detection ability of our model. On the other hand, regarding the classification process, the benign samples are not needed, as in such procedures a classification model only has to decide the family in which a sample, already distinguished as malicious, belongs to.

\noindent \textbf{Graph Structures.} The major target of our approach is to utilize the graphs that depict the temporal evolution of our produced graphs GrG and CvG (i.e., GrTG and CvTG, respectively) in order to measure the graph similarity among test sample and samples that have been already detected as malicious, leveraging their structural modification that take place during the execution time of the programs that they represent. In our work we have a theoretically stable intuition that the factor of time, regarding the structural evolution of a graph is a strong qualitative characteristic that could definitely distinguish the behavior of a program and further be utilized to the development of more elaborated detection and classification techniques over unknown samples.

To this end, we ought to notice that regarding the time quantization procedure, where the time slots where the graph instances have to be retained, there could be applied several different approaches, that would affect the application results. In other words, the implementation of our proposed model on a fine-grained time quantization scheme, would be more precise against a more coarse-grained once, where on the other hand a trade-off between the precision on temporal structural modifications and the construction of more distinguishing patterns poses the basis of further tuning issues.

\vspace*{0.05in}
\noindent \textbf{Similarity Metrics.} Concerning the type of the temporal graph utilized to model the temporal evolution of the corresponding GrG and CvG graphs, (i.e., $T(\widehat{G})$ and $T(G^*)$, respectively) we could deploy two different similarity metrics, namely, $\Delta$-similarity and Cover-similarity metrics, respectively \cite{NiPo2,MpNiPo}, regardless of the applied constructional approach, concerning the discrete or cumulative modification temporal graphs. Next we briefly discuss the two similarity metrics.

\begin{enumerate}[i.]
\renewcommand{\labelitemi}{\scriptsize$\circ$}
\item {\bf $\Delta$-similarity Metric.} The $\delta$-distance, that utilizing the Euclidean distance measures the structural likeness between two given Temporal Graphs, let $T(G_1)$ and $T(G_2)$ denoting with $d_{in}(x)$ (resp. $d_{out}(x)$) is the in-degree (resp. out-degree) of node $x$, and $w_{in}(x)$ (resp. $w_{out}(x)$) is the the average weight of in-coming (resp. out-going) edges of node $x$; is defined as follows.

    \begin{equation}\label{euclidean}
    \delta(T(G_1),T(G_2))=\sum_{i=1}^{k} [\alpha \cdot E_{in}(v_i,u_i) + \beta \cdot  E_{out}(v_i,u_i)]
    \end{equation}

    \vspace*{-0.1in}
    \noindent where,

    $\alpha + \beta = 1$,

    \vspace*{0.1in}

    $E_{in}(v_i,u_i) = \sqrt{(d_{in}(v_i)-d_{in}(u_i))^2 +(w_{in}(v_i)-w_{in}(u_i))^{2^{\phantom{x}}}}$,

    \vspace*{0.1in}

    $E_{out}(v_i,u_i) = \sqrt{(d_{out}(v_i)-d_{out}(u_i))^2 +(w_{out}(v_i)-w_{out}(u_i))^{2^{\phantom{x}}}}$,

%
%
%

    \vspace*{0.1in}

    Since $\Delta$-similarity utilizes the $\delta$-distance metric, we compute it as follows:

    \vspace*{-0.1in}
    \begin{equation}\label{delta}
    \Delta(T(G_1),T(G_2))=\frac{\Gamma}{\Gamma + \delta(T(G_1),T(G_2))}
    \end{equation}

    \vspace*{0.1in}
    \noindent where, $\Gamma \in N^+$ and $0 \leq \Delta() \leq 1$.

\item {\bf Cover-similarity Metric.} In order to perform the malware detection process, we actually compute graph similarity between pairs of Temporal Graphs, let $T(G_1)$ and $T(G_2)$. The graph similarity is computed over the edge set of each pair of Temporal Graphs using the Jaccard similarity metric.

    Let $T(G_1)$ be the Temporal graph of an unknown sample and $T(G_2)$ be the Temporal Graph of a known malware sample. The Cover-Similarity $CS(\cdot)$ of the graphs $T(G_1)$ and $T(G_2)$ is computed as follows:

    \vspace*{0.04in}
    \begin{equation}\label{euclidean}
    CS(T(G_1),T(G_2))=\dfrac{E(T(G_1))\cap E(T(G_2))}{E(T(G_1))\cup E(T(G_2))},
    \end{equation}
    \vspace*{0.00in}

    \noindent where $E(T(G_1))$ and $E(T(G_2))$ are the edge sets of the two tested Temporal Graphs; note that, the vertices $v_1, v_2, \ldots, v_n$ of the graph $T(G_1)$ correspond to the vertices $u_1, u_2, \ldots, u_n$ of the graph $T(G_2)$.
\end{enumerate}

\begin{figure}[t!]
    \hrule\medskip\smallskip
    \centering
    \includegraphics[scale=0.6]{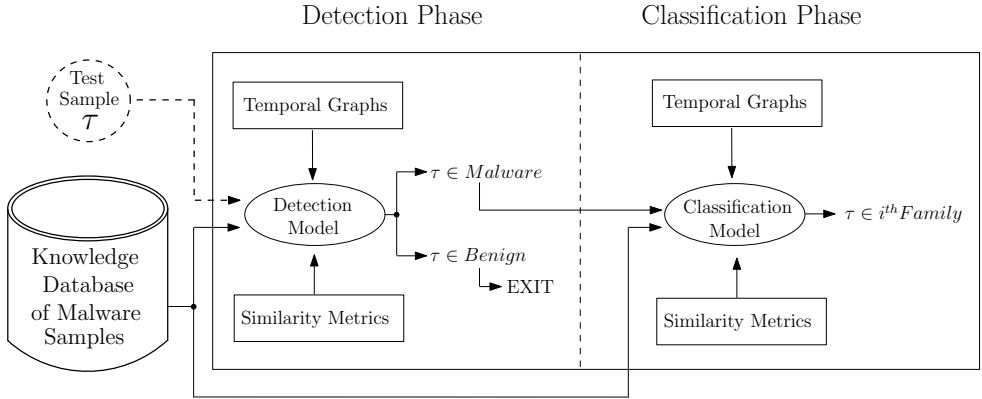}
    \centering
    \smallskip\medskip\hrule\medskip
    \caption{\small{The deployment of malware detection and malware classification processes in our model.}}
\label{fig:fig7}
\end{figure}

\vspace*{0.05in}
\noindent \textbf{Component Deployment.}  The deployment model of the proposed components is consisted by the graph structures utilized to represent the software's behavioral characteristics (i.e., the Temporal Graphs that represent their temporal evolution through time), the knowledge base, that is a database storing Temporal Graphs representing known malware samples, and the similarity metrics developed to capture structural and qualitative commonalities among such behavioral graphs. Our proposed graph-based malware detection and classification model is partitioned into two phases. The first phase concerns the detection procedure, where an unknown sample, let $\tau$, is needed to be detected as malicious or benign. Our mode's implementation utilizes the Temporal Graphs taken from a database of known malware samples and the Temporal Graph of test sample $\tau$ in order to compute their structural similarities across their temporal evolution. The second phase concerns the classification procedure, where an unknown sample, let $\tau$, that has been already detected as malicious is needed to be classified to one of a set of known malware families. Our mode's implementation utilizes the Temporal Graphs taken from a database of known malware samples already been classified to malware families and the Temporal Graph of test sample $\tau$ in order to compute their structural similarities across their temporal evolution and further classify $\tau$ to one malware family from our data-set. In Figure~\ref{fig:fig7}, we represent an abstract overview of the deployment of our proposed graph-based model for malware detection and classification.

\vspace*{0.1in}
\subsection{Malware Detection}

\vspace*{0.05in}
\noindent Next we discuss the operation of our proposed graph based malware detection model, and present an overview on its constructional principles alongside with a brief discussions over its implementation aspects. \\

\vspace*{0.05in}
\noindent \textbf{Model Overview.}We implement our malware detection model by first performing a transformation to the initial ScD graphs, converting them to GrG graphs and CvG graphs respectively, and then we compute for these graphs their corresponding Temporal Graphs (i.e., $T^f(\widehat{G}), T^F(\widehat{G})$ and $T^f(G^*), T^F(G^*)$) as we described in the previous section, and then, for any given test sample we follow the same procedure as to conclude with the computation of $\Delta$-similarity and Cover-similarity metrics in order to measure the structural similarities between the graphs of two samples.

\begin{figure}[t!]
    \hrule\medskip\smallskip
    \centering
    \includegraphics[scale=0.60]{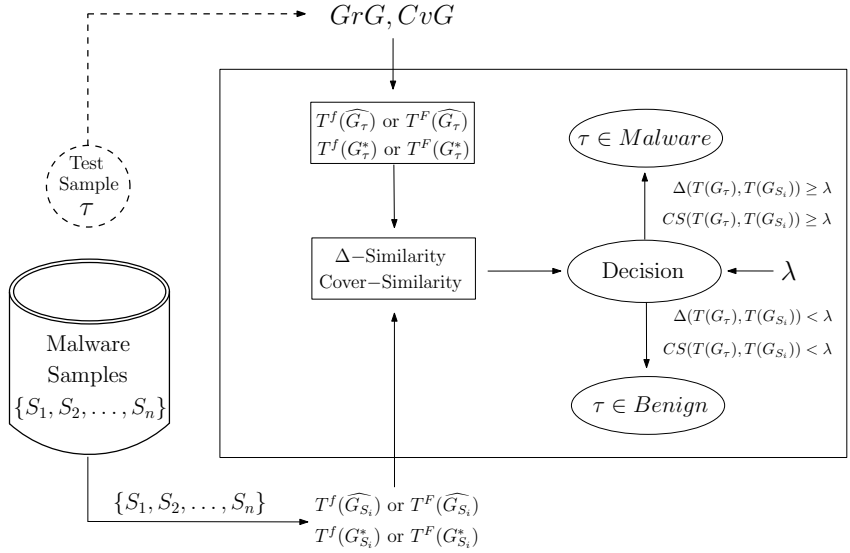}
    \centering
    \smallskip\medskip\hrule\medskip
    \caption{\small{Architecture of the detection model.}}
\label{fig:fig8}
\end{figure}

Next, we describe the main process of determining if an unknown sample is malicious or benign based on the results of our similarity metrics when applied on the corresponding Temporal Graph of a test sample and a set of Temporal Graphs that represent known malicious software samples. In Figure~\ref{fig:fig8} we depict the total architecture of our proposed model for detecting malicious software samples.

\vspace*{0.05in}
\noindent \textbf{Implementation Aspects.} \\
In the example of Figure~\ref{fig:fig8} we suppose we are given an unknown test sample $\tau$ that we do not know if it is malicious, and we are asked to decide if $\tau$ is malicious or benign. Having a database with the Temporal Graphs of known malware samples. Once the corresponding Temporal Graphs have been constructed, we compute the our similarity metrics between the Temporal Graph of $\tau$ and each Temporal Graph that represents a malware sample in our database. So, let $S$ the total number of malware samples in our database, we result to $S$ values in our measurements on our similarity metrics (one per pair $\tau-S_i$), where if the maximum value exhibited is above a predefined threshold $\lambda$ it indicates that $\tau$ is malicious.

\vspace*{0.1in}
\subsection{Malware Classification}

\vspace*{0.05in}
\noindent  Next we discuss the operation of our proposed graph based malware classification model, and present an overview on its constructional principles alongside with a brief discussions over its implementation aspects. \\

\vspace*{0.05in}
\noindent \textbf{Model Overview.}
Our proposed method is based on application our proposed similarity metrics over the set of known malware families in order to classify on them an unclassified malware sample, let $/tau$. More precisely, our method selects the family that is most similar to $\tau$ according to the similarity results exhibited by the measurement of $\Delta$-Similarity and Cover-Similarity metrics, calling that family {\tt dominant} family. More precisely, using our proposed similarity metrics, we iterate over all the members of all the known malware families measuring the similarity between each pair of $\tau$, $M_{ik}$, where $M_{ik}$ is the $i^{th}$ member of the $k^{th}$ malware family. Then, for each family we select a member that is the most similar to $/tau$, according to $\Delta$-Similarity and Cover-Similarity metrics, and denote this member as {\tt representative sample} for this specific family. Finally, among all the {\tt representative samples} for all the known malware families, we select to classify the unclassified test sample $\tau$ to the malware family that its {\tt representative sample} exhibits the maximum similarity with $\tau$ according to $\Delta$-Similarity and Cover-Similarity metrics, denoting this family as {\tt dominant family}.

\begin{figure}[t!]
    \hrule\vspace*{0.3cm}
    \centering
    \includegraphics[scale=0.65]{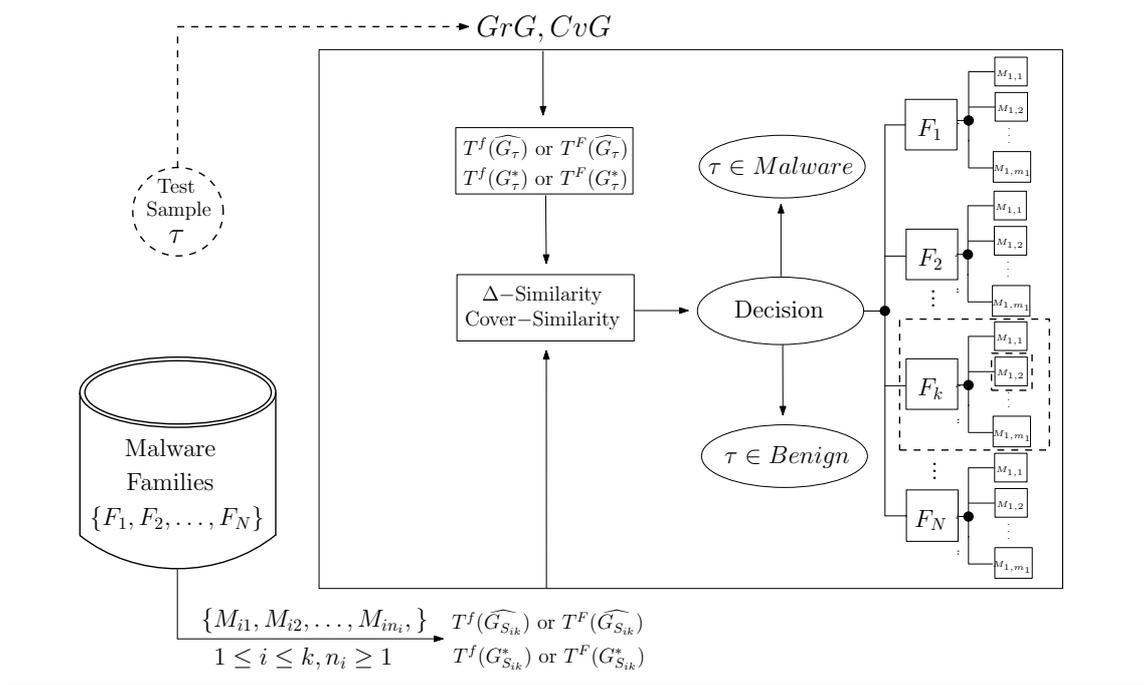}
    \centering
    \medskip\hrule\medskip
    \caption{\small{Architecture of the classification model.}}
    \label{fig:fig9}
\end{figure}

\vspace*{0.05in}
\noindent \textbf{Implementation Aspects.} In the example of Figure \ref{fig:fig9} we show a representation of the procedure for classifying an unknown test sample $\tau$ to a known malware family utilizing the aforementioned methods (i.e., $\Delta$-Similarity and Cover-Similarity metrics). More formally, our classification technique proceeds as follows: given a set of known malware families $F_1, F_2, \ldots, F_N \rbrace $ and an unclassified malware sample $\tau$, we measure the $\Delta$-Similarity and Cover-Similarity metrics over all the members of each family, keeping the maximum result (i.e., {\tt representative sample}) for each family resulting to $N$ results (i.e., $N$ {\tt representative samples}), one per family. Then, we classify the test sample to the family that exhibited the maximum value among all results. In other words, we compute the aforementioned similarity metrics between $\tau$ and all the malware families of the data-set, selecting as the {\tt dominant family}, the one that has the {\tt representative sample} that exhibits the maximum value in our similarity measurements.

\vspace*{0.2in}
\section{Conclusion}
\label{sec:Conclusion}
\vspace*{0.1in}

\noindent In this section we discuss our future work, concerning the implementation of our graph based model for malware detection and classification, and the performance of a series of experiments on our data-set in order to prove the potentials regarding the detection ability and the classification accuracy of our model against malicious samples. Finally we conclude our paper with the remarks regarding mostly the implementation aspects of our work and the potentials our our model expecting the experimental results.

\vspace*{0.1in}
\subsection{Further Research}

\vspace*{0.05in}
\noindent Next, we briefly discuss the potentials and the future work concerning the experimental evaluation of our model regarding its potentials as also drawbacks and limitations that may arise during its implementation.

\vspace*{0.05in}
\noindent \textbf{Potentials.} Several modeling alternates have been arise during the theoretical construction of our graph-based proposed model regarding the temporal evolution of behavioral graphs that represent software samples, regarding their structural modification during time. Our approaches that we discuss briefly next, mostly concern the representation of the structural modifications on the GrG and CvG graphs during time, and how they could also be represented with other structures that do not cooperate graphs, and consequently deserve the application of different manipulation methods.

In the first alternate approach, we could denote the structural evolution of a given by plotting by a discrete distribution of the addition of edges over the graph on specific time buckets (i.e., similar to {\tt epochs}) and create patterns that could be utilized in order to perform pattern-matching over the plot of any given pair of samples (i.e., test and known malware sample). These plots should be construct for the temporal evolution of each corresponding edge pair of two given graphs in order for the patterns to be comparable.

On the other hand, in the second approach of our model, we need to simulate the structural modification of a given graph during time (i.e., temporal evolution of the graph). Similarly to our approach, rather than constructing several graph instances equal to the number of the defined {\tt epochs} and structurally relevant to the applied method regarding the discrete or cumulative modification approach, we could also represent these structural modification (i.e., addition of edges) over the time for each edge (i.e., edge on either GrG or CvG). More precisely, we could define a binary sequence for each edge, where $0$ denotes absence and $1$ denote adition of this edge on the overall graph, and the length of the sequence equals the size of the ScDG (i.e., System-call Depenency Graph). Then, various alignment algorithms could be adopted in order to retrieve similarity patterns among any pair of such sequences, that represent corresponding edges on the graphs of the test and the known malicious samples.

\vspace*{0.05in}
\noindent \textbf{Limitations.} Our proposed graph-based model for malware detection and classification using temporal graphs, despite its theoretical basis, has also some limitations concerning any implementation drawbacks that may arise. The main issue encountered regarding the implementation design concerns the spatial complexity of our approach. More precisely, defining a fine-grained or a coarse-grained quantization of time (i.e., number of {\tt epochs}) would affect to a great extent the space required to store the corresponding Temporal Graph instances. AS easily someone can understand, an implementation of our proposed model on a fine-grained time quantization scheme, would be more precise against a more coarse-grained once. Additionally, further tuning issues arise over the trade-off between the precision on temporal structural modifications and the construction of more distinguishing patterns. However, more sophisticated approaches, such an implementation that utilizes the maximum length of a binary tree in order to bound the quantization would lead to a more stable, rational, effective and efficient approach.

\vspace*{0.1in}
\subsection{Remarks}

\vspace*{0.05in}
\noindent In this paper we designed and presented a graph-based model for malware detection and classification based on relation of Group Relation Graphs and their structural evolution during time (i.e., temporal evolution). On this aspect, to satisfy such demands we proposed the construction of their corresponding Temporal Graphs. Temporal Graphs represent the structural evolution of a graph over the quantum parts of time (i.e., time-slots of on the corresponding time-line that depicts execution-time), that we call {\tt epochs}. The proposed graph-based model for malware detection and classification is organized into two phases, namely detection phase and classification phase, respectively. In order to distinguish malicious from benign samples, and further classify a sample that has been detected as malicious to a malware family from a set of known malware families, we propose the utilization of similarity metrics that measure the graph similarity taking into account structural commonalities among graphs (i.e., $\Delta$-Similarity and Cover-Similarity). Presenting the design principals of our model we discussed its potentials concerning its detection and classification abilities as also minor drawbacks that may be encountered during its implementation in the future.


\end{document}